\documentclass[epj]{svjour}
\usepackage{latexsym}
\usepackage{amssymb}
\usepackage{amsmath}
\usepackage{graphics}
\begin{document}
\title{Shear induced symmetry breaking dynamical states}
\author{K. Premalatha\inst{1}, V K Chandrasekar\inst{2}, L. Senthilkumar\inst{1}, and M. Lakshmanan\inst{3}}

%                  % Do not remove
%
%%\offprints{}          % Insert a name or remove this line
%
\institute{$^{1}$Department of Physics, Bharathiar University, Coimbatore - 641 046, Tamilnadu, India.\\
$^{2}$Department of Physics, Centre for Nonlinear Science \& Engineering, School of Electrical \& Electronics Engineering, SASTRA Deemed University, Thanjavur 613 401, Tamilnadu, India.\\
$^{3}$Department of Nonlinear Dynamics, School of Physics, Bharathidasan University, Tiruchirappalli - 620 024, Tamilnadu, India.
\email{*snkpremalatha@gmail.com}}
\date{Received: ..... / Revised version: .....}
\abstract{
We examine how shear influences the emergence of symmetry-breaking dynamical states in a globally coupled Stuart-Landau (SL) oscillator system with combined attractive and repulsive interactions.  In the absence of the shear parameter, the system exhibits synchronization, nontrivial oscillation death states, and oscillation death states.  However, with the introduction of the shear parameter, we observe diverse dynamical patterns, including amplitude clusters, solitary states, complete synchronization, and nontrivial oscillation death states when the repulsive interaction is weak.  As the strength of the repulsive interaction increases, the system becomes more heterogeneous, resulting in imperfect solitary states.  We also validate the analytical stability condition for the oscillation death region and compare it with the numerical boundary, finding a close match.  Furthermore, we discover that the presence of shear leads to the emergence of symmetry-breaking dynamical states, specifically inhomogeneous oscillation death states and oscillatory cluster states under nonlocal coupling interaction. 
\PACS{
      {PACS-key}{discribing text of that key}   \and
      {PACS-key}{discribing text of that key}
     } % end of PACS codes
} %end of abstract
\maketitle
\section{Introduction}
\label{intro}
\par Nonlinear dynamical systems are prevalent in various natural and artificial systems, spanning disciplines such as earth sciences, sociology, economics, network science, neuroscience and engineering.  The presence of nonlinearities in these systems can give rise to a wide range of rich rhythmic patterns, including turbulence, chaos, and self-sustained oscillations.  Coupled nonlinear oscillators are fundamental models used to describe numerous dynamic behaviors and patterns that mimic real-world phenomena \cite{1,2,3,4,5}.  Due to the interactions between the constituent elements, their individual dynamics and structure of the coupling and coupling topology, an extensive array of dynamical phenomena can be observed. These phenomena encompass various resonances \cite{re1,re2}, synchronization \cite{a,aa}, suppression of oscillations \cite{b}, chimera states \cite{6,7,8,8a,8b}, various chimera death states \cite{38cp,con,hn} and more.
\par While the literature has extensively focused on attractive coupling to explore complex patterns \cite{9,10,11,12}, studies on repulsive coupling, despite its presence in real-world systems, have been relatively scarce \cite{13,14,15}.   Generally attractive or positive interactions drive in-phase synchronization among oscillators whereas repulsive or negative coupling induces out-of phase synchronization.   As a result, recent investigations have increasingly concentrated on mixed attractive and repulsive couplings due to their relevance in various realistic systems, such as biological \cite{16}, chemical \cite{17,18}, sociological\cite{19}, ecological \cite{20,21} and socio-technical systems \cite{22,23,24}.  In Ref. \cite{25} many hysteresis loops and limit cycle activity were noticed in excitatory and inhibitory couplings among sub-population of model neurons.  Nima et al.\cite{26} examined the dynamics of excitation and inhibition in the neocortex of humans and monkeys, reporting balanced activity of ensembles in all states of wake-sleep cycle.  Furthermore, Andrea Giron et al. \cite{20} investigated the role of species in their organization and specifically described the ecological equilibrium between facilitation and competition in plant communities.  
\par Moreover, attractive and repulsive couplings give rise to a wide range of symmetry breaking dynamical phenomena, including chimera states \cite{29}, solitary states \cite{27} and oscillation death states \cite{27b}.   For instance, when attractive and repulsive interactions are present, the system's symmetry is disrupted, leading to the existence of chimera-like states in a globally coupled Kuramoto-Sakaguchi system of oscillators \cite{27a}. In Ref. \cite{27}, Maistrenko et. al. discovered that a transition from coherence to incoherence occurs through solitary states at the edge of synchrony.  They also demonstrated that solitary states emerge when attractive and repulsive interactions act in anti-phase.  Furthermore, the existence of solitary states was observed in a system of coupled FitzHugh-Nagumo oscillators, where attractive inter-layer and repulsive intra-layer interactions were present \cite{30}.  Nonisochronicity (also known as shear) plays a significant role in inducing symmetry-breaking dynamics and is responsible for the emergence of chimera states in a globally coupled array of complex Ginzburg-Landau oscillators \cite{32b} and van der Pol oscillators \cite{32a}.  In Ref. \cite{8}, a diverse range of dynamical states and the consequences of symmetry breaking in coupling were reported in a network of globally coupled Stuart-Landau oscillators with nonisochronicity.  Recently, the symmetry breaking dynamics resulting from attractive and repulsive interactions were observed in two directly coupled Stuart Landau oscillators in the absence of nonisochronicity \cite{32}.  Motivated by these findings, our goal here is to investigate the effect of shear/nonisochronicity in inducing symmetry breaking dynamical states in a network of globally coupled Stuart-Landau oscillators under combined attractive and repulsive interactions.
\par  Based on the aforementioned facts, our objective is to investigate the influence of shear in inducing a variety of dynamic transitions in a network of globally coupled Stuart-Landau oscillators. These oscillators will be subject to combined asymmetric attractive and repulsive interactions, which break the system's symmetry. In the absence of shear, the system exhibits only fully synchronized states, nontrivial oscillation death states, and oscillation death states. However, by introducing shear into the system, we observe the emergence of various collective dynamic states, including amplitude cluster states, solitary states, complete synchronization, nontrivial oscillation death states, and oscillation death/chimera death states, particularly for weak repulsive coupling. The system's transition to oscillation death states depends on the strength of the repulsive interaction, and the presence of strong repulsive coupling leads to increased asymmetry in the dynamical region, resulting in the existence of imperfect solitary states. We also explore the analytical stability of the oscillation death region, which closely aligns with the numerical results. Additionally, we examine the impact of shear on the system's dynamic states under nonlocal coupling interaction and observe the occurrence of symmetry-breaking dynamical states, such as inhomogeneous oscillation states and oscillatory cluster states, for sufficiently high shear values.
\par The structure of the paper is as follows: Section II provides a description of the model equation. In Section III, we discuss the dynamic regimes in the absence of shear in a globally coupled system of Stuart-Landau oscillators with asymmetric attractive and repulsive couplings. Section IV examines the consequences of introducing shear with asymmetric attractive and repulsive couplings.  Section-V deals with the impact of introducing shear under nonlocal coupling with combined attractive and repulsive couplings. Finally, the results are summarized in the concluding section.
\section{Model}
 The globally coupled array of Stuart-Landau oscillators with combined attractive and repulsive interactions is given by the following set of equations \cite{33,33a,33b},
 \begin{eqnarray}
\dot{x_j}=x_j-\omega y_j-(x_j^2+y_j^2)(x_j+cy_j)+\frac{\varepsilon_1}{N} \sum_{k=1}^{N} (x_k-x_{j}), \nonumber\\
\dot{y_j}=\omega x_j+y_j-(x_j^2+y_j^2)(y_j-cx_j)-\frac{\varepsilon_2}{N} \sum_{k=1}^{N} (y_k-y_{j}),
\label{eq1}
\end{eqnarray}
where $j$ runs from $1,2,3....,N$.  The frequency parameter $\omega$ is real.  $c$ is the shear (or) nonisochronicity parameter which indicates the dependence of the angular velocity on the radial co-ordinate.  Here $\varepsilon_1$ and $\varepsilon_2$  are the attractive and repulsive coupling strengths.  If $\varepsilon_1=\varepsilon_2$ in Eq. (\ref{eq1}), it corresponds to the case of oscillators being coupled with symmetric interaction while $\varepsilon_1 \neq \varepsilon_2$ corresponds to asymmetric interaction strengths.  By solving Eq. (\ref{eq1}) numerically using the fourth order Runge-Kutta method, we analyze the results.  In our simulations, we choose the number of oscillators $N$ to be equal to 100 and $\omega=1$ to numerically solve Eq. (\ref{eq1}).  The initial state of the oscillators is uniformly distributed between -1 and +1. We have verified that the results are independent of increasing the total number of oscillators.
\par In Ref. \cite{8}, the authors have considered Eq. (\ref{eq1}) for $\varepsilon_2=0$ and identified a rich variety of dynamical states with respect to the shear strength.  In addition, shear in the system plays an important role in controlling synchronization in a network of nonidentical oscillators.  Interplay between shear and natural frequency of the system can cause an anomalous effect which can either inhibit or enhance synchronization \cite{33}. In the present study, we analyze the consequences of shear, $c$, under attractive and repulsive interactions and the results are clearly explained with two parameter diagrams in the following sections. 
\section{Impact of attractive and repulsive couplings in the absence of shear}
\par
\begin{figure*}[ht]
\begin{center}
\resizebox{0.60\textwidth}{!}{\includegraphics{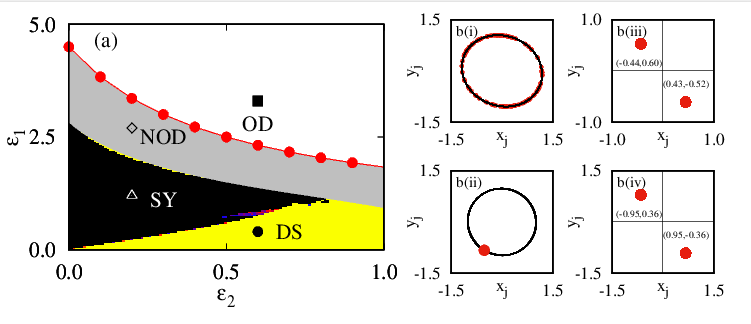}}\\
\end{center}
\caption{ (a) Two parameter phase diagram in the parametric space ($\varepsilon_2,\varepsilon_1$) in the absence of shear with $\omega=1$.  Snapshots of the variable in the complex plane $z_j$: b(i) Desynchronized state, b(ii) synchronized state, b(iii) nontrivial oscillation death state and b(iv) oscillation death state.  Trajectories are shown by black solid lines and red dots represent the snapshots of the variables $z_j$ in the complex plane.  Here $\bullet$, $\triangle$, $\Diamond$ and $\blacksquare$ mark the parameter values corresponding to desynchronized state, synchronized state, nontrivial oscillation death state and oscillation death state, respectively.  The region DS is the desynchronized region, $SY$ represents synchronized state, $NOD$ represents the nontrivial oscillation death state and $OD$ represents oscillation death state.  Analytical boundary is represented using red/grey color dots.  These figures represent the existence of different dynamical regimes in the absence of shear.} 
\label{fig1a}
\end{figure*}

 In order to facilitate the study of different dynamical regimes of the system in the absence of shear, we plotted a two parameter diagram in the parametric space ($\varepsilon_2,\varepsilon_1$) in Fig. \ref{fig1a}(a).  Here the repulsive interaction range is considered in the region $0.0<\varepsilon_2<0.83$, and the system follows the transition route from desynchronized state (DS) to oscillation death state (OD) through synchronized state (SY) and nontrivial oscillation death state (NOD) as the attractive interaction strength $\varepsilon_1$ is varied between $0$ and $5$.  By strengthening the interaction beyond $\varepsilon_2>0.83$, the system attains oscillation death states from desynchronized state through nontrivial oscillation death states with respect to $\varepsilon_1$.  More details about the dynamical regimes are illustrated with the evolution of the dynamical variables ($x_j,y_j$) (solid line) with their snapshots (dots) in the complex plane ($x_j,y_j$) in Figs. \ref{fig1a}b(i)-(iv).  In the set of Figs. \ref{fig1a}(b), incoherent behaviour of the system of oscillators is given in Fig. \ref{fig1a}(a) for the value of ($\varepsilon_2,\varepsilon_1) \in (0.6,0.4)$.  Figure \ref{fig1a}(b) shows the synchronization among the oscillators (($\varepsilon_2,\varepsilon_1) \in (0.2,1.2)$).  Distribution of nontrivial oscillation death (NOD) state is plotted in Fig. \ref{fig1a}(c) for ($\varepsilon_2,\varepsilon_1) \in (0.2,2.7)$.  Nontrivial oscillation death represents the manifestation of two branches of the inhomegeneous steady states appearing at $x_1$ and $x_2\ne x_1$, (ie., $x_2=x_1\pm\delta$)  in contrast to oscillation death state as illustrated in Fig. \ref{fig1a}(d) for ($\varepsilon_2,\varepsilon_1) \in (0.6,2.9)$.  Here $x_1$ belongs to upper branch of the inhomohgeneous steady state and $x_2$ corresponds to the lower branch of the inhomogeneous steady state and the value of $\delta$ ($=|x_2-x_1|$) is small.  Here the steady state values are (-0.44,0.63) and (0.43,-0.52).   Figure \ref{fig1a}(d) illustrates the manifestation of the two branches of inhomogeneous steady states occurring at $x_1$ and $x_2=-x_1$. Here the steady state values are (-0.95,0.36) and (0.95,-0.36).  Oscillation death region is a multi stable region and it coexists with synchronized state and chimera death state for appropriate choice of the initial condition.  
\par On the other hand, the oscillation death state represents the situation where the total population is split into two groups of inhomogeneous steady states.  Hence the dynamical equations for the two populations of oscillators can be written in the form as 
\begin{eqnarray}
\frac{dx_{h1}}{dt}&=&f(x_{h1})+\varepsilon_1[p x_{h1}+q x_{h2}-x_{h1}],\nonumber\\
\frac{dy_{h1}}{dt}&=&f(y_{h1})-\varepsilon_2 [p y_{h1}+q y_{h2}-y_{h1}],\nonumber\\
\frac{dx_{h2}}{dt}&=&f(x_{h2})+\varepsilon_1[p x_{h2}+q x_{h1}-x_{h2}],\nonumber\\
\frac{dy_{h2}}{dt}&=&f(y_{h2})-\varepsilon_2 [p y_{h2}+q y_{h1}-y_{h2}],
\label{eq2}
\end{eqnarray}
where $f(x_{h1})=x_{h1}-\omega y_{h1}-(x_{h1}^2+y_{h1}^2)(x_{h1}+cy_{h1})$, $f(x_{h2})=x_{h2}-\omega y_{h2}-(x_{h2}^2+y_{h2}^2)(x_{h2}+cy_{h2})$, $f(y_{h1})=\omega x_{h1}+y_{h1}-(x_{h1}^2+y_{h1}^2)(y_{h1}-cx_{h1})$, $f(y_{h2})=\omega x_{h2}+y_{h2}-(x_{h2}^2+y_{h2}^2)(y_{h2}-cx_{h2})$.  Here $(x_{h1},y_{h1})$ and $(x_{h2},y_{h2})$ are the states of the oscillators corressponding to two groups of (lower and upper) inhomogeneous steady states and they are having $p$ and $q=1-p$ number of oscillators, respectively. Note that the total number of oscillators gets equally split and occupies lower and upper inhomogeneous steady states.  Hence we consider the value of $p$ as $0.5$ (according to numerical results).  The above system has the trivial fixed point at ($x_1^*,x_2^*,y_1^*,y_2^*$)=(0,0,0,0) and the nontrivial fixed points can be expressed as 
\begin{eqnarray}
x_{1,2}^*&=&\pm(\frac{\varepsilon_1+\varepsilon_2-\gamma_0)}{\sqrt{2}c(\varepsilon_1-1)-\omega}\sqrt{\frac{c^2\gamma_0\varepsilon_1^2+\gamma_1\varepsilon_1+\gamma_2}{(1+c^2)^2(\varepsilon_1+\varepsilon_2)}},\\
\label{eq3}
y_{1,2}^*&=&\frac{c(\varepsilon_1-1)-\omega}{(\varepsilon_2-\varepsilon_1)-\omega}x_{1,2}^*
\label{eq4}
\end{eqnarray}
with $\gamma_0=\varepsilon_1^2((\varepsilon_1+\varepsilon_2)^2+2 c^2\varepsilon_2)+4(c+\omega)(\varepsilon_1-\varepsilon_2+c+\omega)$, $\gamma_1=1+4c^2+2\omega c(1-c^2)+\varepsilon_2+3c^2\varepsilon$ and $\gamma_2=\varepsilon_2(1-3c^2-4c\omega)+(1-c^2-2c\omega+\varepsilon_2)+2c^2(c\omega-2)-6c\omega-2\omega^2(1-c^2)$.   These fixed points break the symmetry of the system which causes the stabilization of the system into two inhomogeneous steady states.  We find the stability condition for oscillation death states by using linear stability analysis.  The corresponding eigenvalues of the system (\ref{eq2}) are
{\begin{eqnarray}
 \lambda_{1,2}&=&\frac{1}{2}(2-a_0-4a_3)\pm\sqrt{b_1},\label{eq5}\\
\lambda_{3,4}&=&1-2a_3\pm\sqrt{b_2},
\label{eq6}
\end{eqnarray}
where $a_0=\varepsilon_1+\varepsilon_2$, $a_1=\varepsilon_1-\varepsilon_2$, $a_3=x^2+y^2$, $a_4=x^2-y^2$, $b_1=a_1^2+4a_3^2+4a_4(a_1-3c^2x^2)-4\omega(4cx^2+y^2(c+3x)+\omega)+8xy(c\varepsilon_2+3xy^2)$ and $b_2=-\omega^2+6x^2y^3+a_3^2-3cx(cxa_4+y^2(3x^2+y(2+y)))-c\omega(4x^2+y^2(1+3x))$.  From the eigenvalue equations (\ref{eq5}) and (\ref{eq6}), the inhomogeneous steady states become linearly stable at 
\begin{equation}
\varepsilon_1=\frac{1}{2(2\varepsilon_2+1)}(1+2\varepsilon_2+c^2+4\omega+\omega^2).
\label{eq7}
\end{equation}
By using the above condition, we plotted the boundary for oscillation death region in Fig. \ref{fig1a}(a) in the absence of shear $(c=0)$.  Thus we can infer that the OD states are stable above the analytical boundary line (red/grey dots).  The analytical boundary line closely matches the numerical results.
\section{Consequence of shear on dynamical states under attractive and repulsive coupling}
\begin{figure*}[ht!]
\begin{center}
\resizebox{0.9\textwidth}{!}{\includegraphics{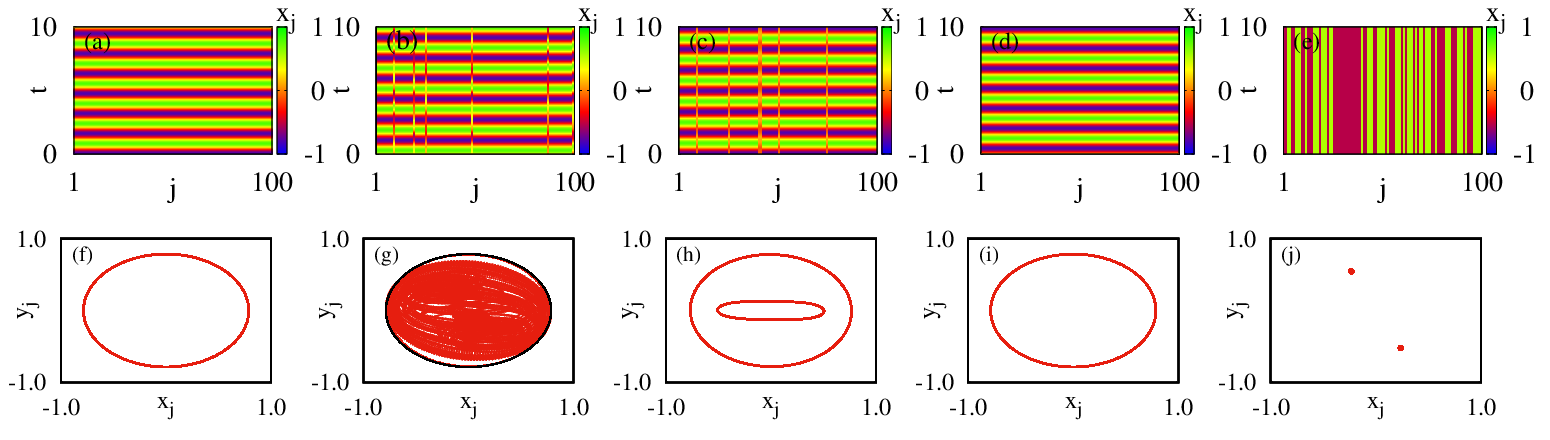}}\\
\end{center}
\caption{Spatiotemporal plots depicting (a) complete synchronization for $\varepsilon_1=0.7$, (b) solitary state for $\varepsilon_1=1.8$, (c) amplitude cluster state for $\varepsilon_1=2.8$, (d) complete synchronization for $\varepsilon_1=3.2$, (e) oscillation death state for $\varepsilon_1=5.9$.  In all the cases we have chosen $c=3$ and $\varepsilon_2=0.5$ and the value of $\omega$ is fixed as $\omega=1$.  Panels (f)-(j) are the associated phase portraits of the oscillators for panels (a)-(e).  Note that the above figures represent the transition of different dynamical states formed by varying the strength of attaractive coupling when the repulsive interaction is fixed as $\varepsilon_2=0.5$.} 
\label{fig1}
\end{figure*} 
\begin{figure*}[ht!]
\begin{center}
\resizebox{0.6\textwidth}{!}{\includegraphics{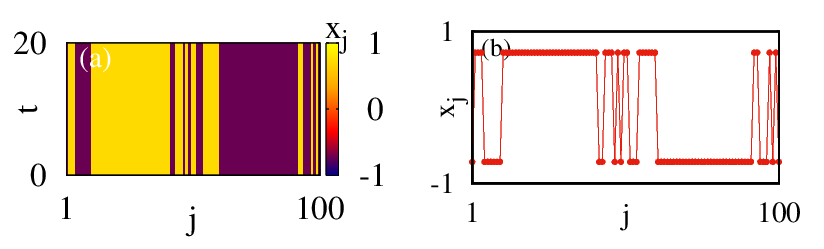}}\\
\end{center}
\caption{Existence of chimera death state for spatially prepared initial condition in the oscillation death region. (a) Spatiotemporal plot and (b) snapshot of the variables $x_j$ for $\varepsilon_1=6.2$ with  $c=3$ and $\varepsilon_2=0.5$. } 
\label{fig1s}
\end{figure*} 
 In order to facilitate the understanding of the various observed results, we first consider that the repulsive interaction between the oscillators as weak and the parameter values are fixed as $\omega=1$ and $\varepsilon_2=0.5$.  The value of shear is chosen as $c=3$.  The reason for such a choice is that only for sufficiently high value of shear solitary states are induced.  Disparate dynamical states are noticed by varying the attractive coupling strength $\varepsilon_1$ in Fig. \ref{fig1}.  Initially for very small values of $\varepsilon_1$, the oscillators in the array exhibit desynchronous motion in time which is not shown here.  On slightly increasing the value of $\varepsilon_1$, we can observe the existence of complete synchronization as depicted in Fig. \ref{fig1}(a) and (f) for $\varepsilon_1=0.7$.  By increasing the strength of the attractive coupling to $\varepsilon_1=1.8$, we find that a few of the oscillators get deviated from the synchronized group as shown in the spatiotemporal plot Fig. \ref{fig1}(b).  Frequencies of such deviated oscillators are randomly distributed while the frequency of the synchronized oscillators is the same which is not illustrated here.  Such split up of individual solitary oscillators from the synchronized group oscillates with different frequencies which give rise to the existence of symmetry breaking dynamical state, namely a solitary state.  Also, it can be noted that the oscillators in the synchronized group are oscillating periodically while the oscillations corresponding to solitary oscillator group are quasi-periodic in nature.  It is depicted by plotting the phase portrait of the oscillators in Fig. \ref{fig1}(g).  Upon increasing the coupling strength further, the system of oscillators attains two groups of synchronized clusters namely amplitude clusters for $\varepsilon_1=2.8$ as in Fig. \ref{fig1}(c).  The phase portrait of the corresponding oscillators is illustrated in Figs. \ref{fig1}(h).  On further increasing the interaction strength $\varepsilon_1=3.2$, the system again attains complete synchronization, Figs. \ref{fig1}(d) and (i).  Upon increasing the attractive coupling strength to $\varepsilon_1=5.9$, the presence of asymmetry in the coupling strengths breaks the symmetry of the system which gives rise to the emergence of nontrivial oscillation death states as in Figs. \ref{fig1}(e) and (j).  Finally, the system gets settled into oscillation death state upon increasing the coupling interaction which is not shown here for simplicity.  In the oscillation death state, oscillators populate coupling dependent stable steady states due to symmetry breaking in the system which gives rise to stable inhomogeneous steady states.  Another important point to be noted here is that for spatially prepared initial conditions, chimera death state emerges in the oscillation death region as in Figs. \ref{fig1s}(a) and \ref{fig1s}(b).  Chimera death states represent the coexistence of spatially coherent (where neighboring oscillators attain the same branches of inhomogeneous steady states) and incoherent (where neighboring oscillators jump among the different branches of inhomogeneous steady states) oscillation death states.
\begin{figure*}[ht!]
\begin{center}
\resizebox{0.9\textwidth}{!}{\includegraphics{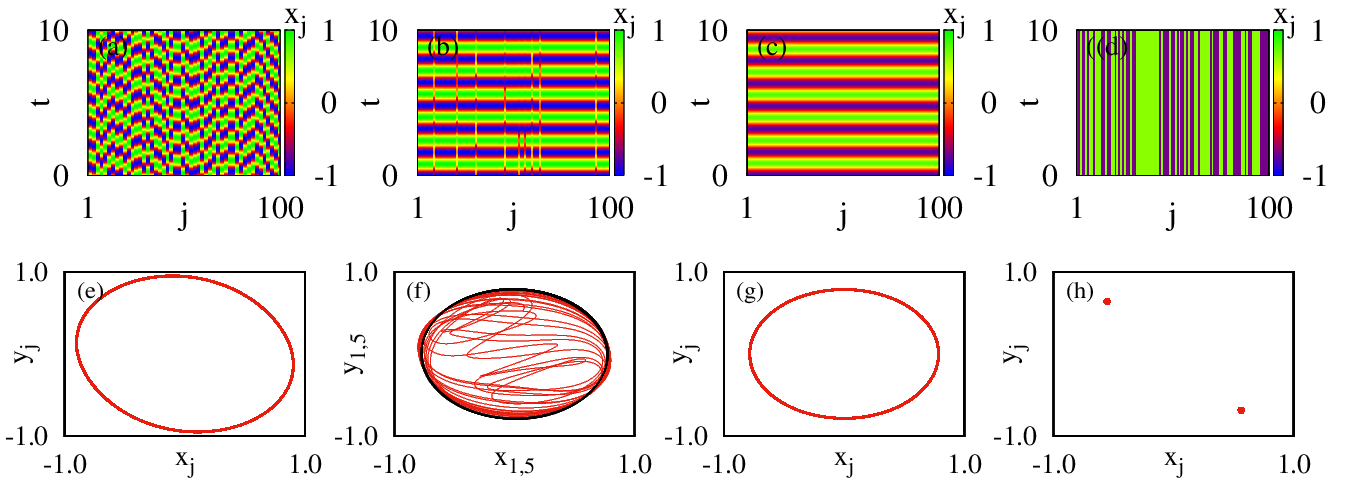}}
\end{center}
\caption{Spatiotemporal plots depicting (a) desynchronization for $\varepsilon_1=0.01$, (b) imperfect solitary state for $\varepsilon_1=2.2$, (c) complete synchronization for $\varepsilon_1=3.0$, (e) oscillation death state for $\varepsilon_1=5.9$.  In all the cases we have chosen $c=3$ and $\varepsilon_2=1.5$.  Panels (e)-(h) represent the phase portraits of the oscillators for panels (a)-(d).  The above figures represent the transition of different dynamical states formed by varying the strength of attractive coupling when repulsive interaction is fixed as $\varepsilon_2=1.5$.} 
\label{fig3}
\end{figure*} 
\begin{figure*}[ht!]
\begin{center}
\resizebox{0.70\textwidth}{!}{\includegraphics{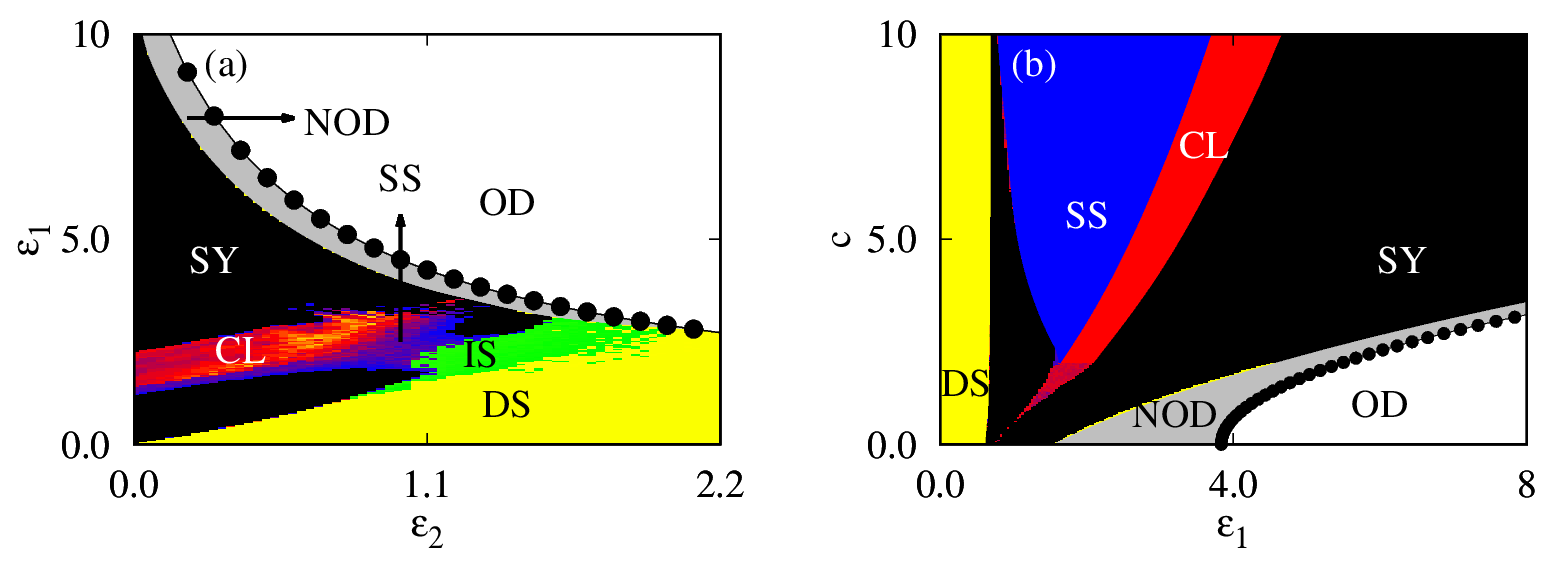}}\\
\end{center}
\caption{Two parameter phase diagram (a) in the parametric space ($\varepsilon_2,\varepsilon_1$) with fixed $c=3$ and (b) in the parametric space ($\varepsilon_1,c$) with fixed $\varepsilon_2=0.5$. The region SY is the synchronized region, $CL$ represents the amplitude cluster state, regions $SS$ and $IS$ represent solitary state and imperfect solitary state, respectively, and regions $OD$ and $NOD$ represent oscillation death state and nontrivial oscillation death state, respectively.  Analytical stability boundary for oscillation death state is plotted using black color dots.  This figure depicts the existence of various dynamical transitions in the presence of shear.} 
\label{fig4}
\end{figure*} 
 
\par Next, we analyze the dynamical behaviour of the system by strengthening the repulsive interaction between the oscillators.  For this purpose, we fix the repulsive interaction strength $\varepsilon_2$ as $\varepsilon_2=1.5$.  By varying the attractive coupling interaction $\varepsilon_1$, the dynamical transitions observed are depicted in Fig. \ref{fig3}.  For sufficiently small values of $\varepsilon_1$, the oscillators are oscillating asynchronously with periodic oscillations with random phases as in Figs. \ref{fig3}(a) and \ref{fig3}(e) in contrast to the case of synchronous oscillation where the oscillators are oscillating with same phases (Fig. \ref{fig1}(a)).  Upon increasing the coupling interaction, there arises an inhomogeneity in the system which leads to the existence of an imperfect solitary state for $\varepsilon_1=2.2$  as in Fig. \ref{fig3}(b).  An imperfect solitary state represents the situation where the deviated solitary oscillators exhibit periodic motion for a certain time after which periodic oscillations of the solitary oscillators become aperiodic in nature.  This tendency repeats irregularly with respect to time.  This behaviour is clearly illustrated by plotting the phase portrait of the two chosen oscillators $z_{1,5}$ in Fig. \ref{fig3}(f).  On further increasing the value $\varepsilon_1$ to $\varepsilon_1=3.0$, the system gets completely synchronized with periodic oscillations (as in Figs. \ref{fig3}(c) and \ref{fig3}(g)).   As the strength of the repulsive coupling interaction increases, the inhomogeneity in dynamical behaviour increases which results in the existence of nontrivial oscillation death states as shown in Fig. \ref{fig3}(d) and \ref{fig3}(h) for $\varepsilon_1=5.9$.  On further strengthening the coupling interaction, the system attains OD state which is not shown here.  We also note here that inclusion of sufficient strengths of shear and repulsive interaction promotes the symmetry breaking dynamics in the system which results in the existence of imperfect solitary states and nontrivial oscillation death states.  Hence we can conclude that an interplay of shear and symmetry breaking interaction leads to the emergence of symmetry breaking dynamical states in the system.    
 \begin{figure*}[ht!]
\begin{center}
\resizebox{0.9\textwidth}{!}{\includegraphics{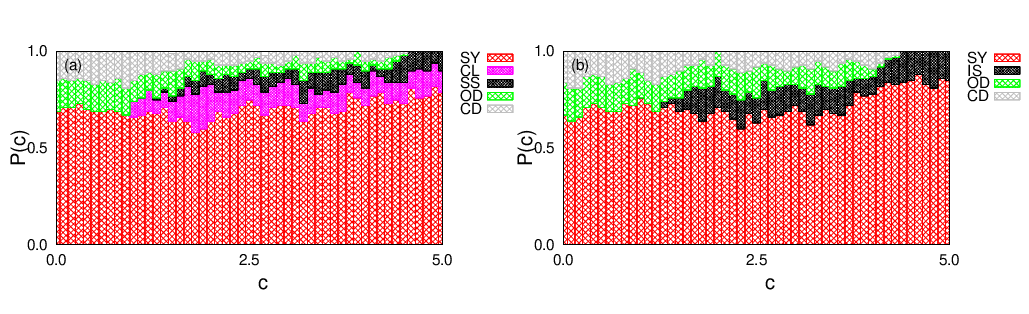}}\\
\end{center}
\caption{Probability of occurrence P(c) for various dynamical states as a function of $c$ for two different $\varepsilon_2$ values: (a) for $\varepsilon_2=0.5$ and (b) for $\varepsilon_2=1.5$.  Various dynamical patterns with colors represent the existence probability of various dynamical states.  Other parameter value: $\varepsilon_1=2.5$.  In this figure, one can infer that the probability of existence of synchronized state is greater than that of the other dynamical states in both the cases of weak and strong coupling interactions.} 
\label{fig5}
\end{figure*}
  \par In order to understand the global picture of the system, a two parameter phase diagram in a wide range of attractive and repulsive interaction strengths ($\varepsilon_2$,~$\varepsilon_1$) in the presence of shear is depicted in Fig. \ref{fig4}(a).  Here we have chosen the shear parameter value as $c=3$.  From this figure, we can observe that depending on the strength of the repulsive coupling, the system exhibits different dynamical transitions with respect to $\varepsilon_1$.   For weak repulsive coupling strength in the range ($0.0<\varepsilon_2<1.18$),  the system exhibits swing of synchronized states through solitary and amplitude cluster states and attains OD via nontrivial OD states.  Then the system transits from desynchronized states to oscillation death states through imperfect solitary states, synchronized states and nontrivial OD in the range $(1.18<\varepsilon_2<1.58)$.  On further increasing $\varepsilon_2$  beyond the value of $1.58$, we find that there exists a transition from desynchronization to OD through imperfect solitary states and nontrivial OD states.  Note that there exists a coexistence of synchronized state with amplitude cluster, solitary and imperfect solitary states in the regions $CL$, $SS$ and $IS$, respectively.  On the other hand, the region $OD$ is the multistability region between synchronized state and oscillation death/chimera death states.  For spatially prepared initial conditions, we can observe the chimera death states whereas the choice of the initial condition near synchronized solution leads to the existence of synchronized state.  In order to know how the dynamical transitions are affected by increasing the strength of the shear, we have plotted the two parameter phase diagram in parametric space ($\varepsilon_1,c$ ) for fixed $\varepsilon_2=0.5$ in Fig. \ref{fig4}(b).  Here we can note that there occurs no change in the behaviour of dynamical transitions and only their dynamical regions are widened with respect to shear.  The important thing to be noted here is that the symmetry breaking dynamical states namely amplitude cluster, solitary states and imperfect solitary states are observed only when the strength of shear is sufficiently large.  Note that the nontrivial OD region is suppressed as the shear value rises, despite the fact that shear causes inhomogeneity (or symmetry breaking) dynamics in the oscillatory domain.  By using the condition given in Eq. (\ref{eq7}), the stability boundary is plotted in Fig. \ref{fig4}(a) with black dots for $c=3$.  One can find that the oscillation death state is stable above the analytically observed boundary.  In Fig. \ref{fig4}(b), the oscillation death state is stable below the analytical boundary for $\varepsilon_2=0.5$.  Hence we can conclude that the analytical results closely match the numerical results.  Various dynamical transitions with respect to repulsive coupling strength ($\varepsilon_2$) is given in Table. 1.
\begin{table}[h!]
\begin{center}
\begin{tabular}{ ||c|c||c|| } 
\hline
Coupling range & Dynamical transitions  \\
\hline
\hline
 0.0  $< \varepsilon_2 < 1.18$ & DS $\rightarrow $SY$\rightarrow$ SS $\rightarrow$ CL $\rightarrow$ SY $\rightarrow$ NOD $\rightarrow$ OD \\
 1.18  $< \varepsilon_2 < 1.58$ & DS $\rightarrow$ IS $\rightarrow$ SY $\rightarrow$ NOD $\rightarrow$ OD \\ 
 1.58  $< \varepsilon_2 < 2.02$ & DS $\rightarrow$ IS $\rightarrow$ NOD $\rightarrow$ OD\\ 
  $\varepsilon_2 > 2.02$ & DS $\rightarrow$  OD \\ 
\hline
\end{tabular}
\caption{Various dynamical transitions were obtained in the case of combined asymmetric attractive and repulsive couplings.  Here the various dynamical states are desynchronized state (DS), synchronized state (SY), cluster state (CL), solitary state (SS), imperfectly solitary state (IS), nontrivial oscillation death state (NOD), oscillation death state (OD).}
\end{center}
\end{table}
 \par The probability of occurrence of different attractors for a distribution of 300 initial states $(x_j,y_j)$ $\in$ (-3,3) as a function of shear $c$ with fixed $\varepsilon_2=0.5$ in Fig. \ref{fig5}(a) and $\varepsilon_2=1.5$ in Fig. \ref{fig5}(b) are shown.  Probability of occurrence can be calculated from the ratio between the number of initial conditions going to different attractors ($A_{ic}$) and total number of initial conditions considered for realizations ($T_{ic}$) and it can be represented as $P(c)=\frac{A_{ic}}{T_{ic}}$. In both the figures, the other parameter value is fixed as $\varepsilon_1=2.5$.  From this figure, we can note that the probability of existence of synchronized state is greater than that of the other dynamical states in both the cases of weak and strong coupling interactions.  Moreover, it may be observed that the probability of emergence of solitary and imperfect solitary states exists for sufficient value of shear strength.  
\section{Impact of shear on dynamical states under nonlocal coupling:}
\par Next, we analyze the impact of shear in the case of nonlocally coupled Stuart-Landau oscillators. Their dynamical equations are given by
 \begin{eqnarray}
%\dot{z_j}=(1+i\omega_{j})z_{j}-(1- ic)|z_{j}|^2 z_{j}+\varepsilon_1 \sum_{k=1}^{N} Re(z_k-z_{j}) -i\varepsilon_2 \sum_{k=1}^{N} Im(z_k-z_{j}),
\dot{x_j}=x_j-\omega y_j-(x_j^2+y_j^2)(x_j+cy_j)+\frac{\varepsilon_1}{2P} \sum_{k=N-P}^{N+P} (x_k-x_{j}), \nonumber\\
\dot{y_j}=\omega x_j+y_j-(x_j^2+y_j^2)(y_j-cx_j)-\frac{\varepsilon_2}{2P} \sum_{k=N-P}^{N+P} (y_k-y_{j}),
\label{eqn}
\end{eqnarray}
where the attractive and repulsive interaction strengths, namely $\varepsilon_1$ and $\varepsilon_2$, respectively control the nonlocal coupling in the system. The coupling range is represented as $r=\frac{P}{N}$, where $N$ is the total number of oscillators and $P$ corresponds to the number of nearest neighbors in both direction. The natural frequency of the oscillator $\omega$ is fixed as $\omega=1$. 

\begin{figure*}[ht!]
\begin{center}
\resizebox{0.5\textwidth}{!}{\includegraphics{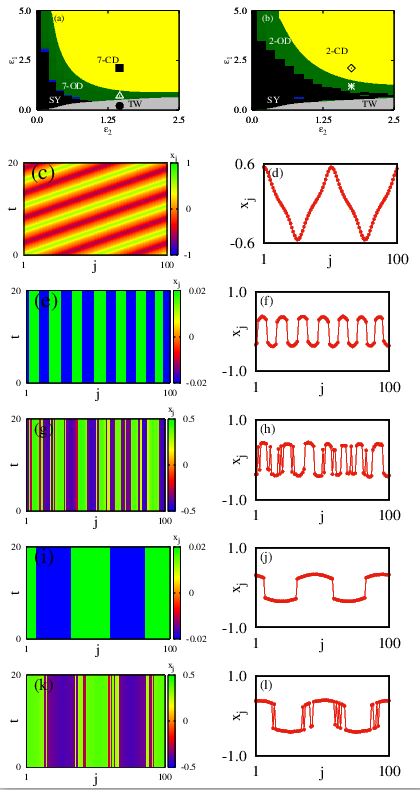}}\\
\end{center}
\caption{Two parameter phase diagram in the parametric space ($\varepsilon_2,\varepsilon_1$) in the absence of shear: (a) for nonlocal coupling range $r=0.1$ and (b) for nonlocal coupling range $r=0.4$.  Space-time plots and their corresponding snapshots of the variables $x_j$ for (c)-(d) traveling wave state, (e)-(f) 7-cluster oscillation death state for $r=0.1$, (g)-(h) 7-cluster chimera death state for $r=0.1$, (i)-(j) 2-cluster oscillation death state for $r=0.4$, (k)-(l) 2-cluster chimera death state for $r=0.4$.  In Figs. (a) and (b), $\bullet$, $\triangle$, $\blacksquare$, $*$ and $\diamond$ mark the parameter values corresponding to traveling wave state, 7-OD, 7-CD, 2-OD and 2-CD states, respectively.  The region SY is the synchronized region, $TW$ represents the traveling wave state, regions $7-OD$ and $7-CD$ represent 7-cluster oscillation death state and 7-cluster chimera death state, respectively and regions $2-OD$ and $2-CD$ represent 2-cluster oscillation death state and 2-cluster chimera death state, respectively. Other parameter values are $\omega=1$ and $c=0$.  Figures (c)-(l) depict the characteristic nature of the dynamical states with respect to nonlocal coupling range in the absence of shear.} 
\label{fig7}
\end{figure*}
\par In this section, we will discuss the impact of shear on dynamical transitions under combined attractive and repulsive interactions under nonlocal coupling.  In order to illustrate the results, in the absence of shear, we plotted the two parameter phase diagram in the parametric space ($\varepsilon_2,\varepsilon_1$) for two different values of coupling range, $r=0.1$ and $r=0.4$, in Fig. \ref{fig7}(a) and \ref{fig7}(b), respectively.  From Fig. \ref{fig7}(a) it is clear that for a small value of nonlocal coupling range $r=0.1$, by varying the coupling strength $\varepsilon_1$ in the range of repulsive interaction $\varepsilon_2$ $\in$ $(0.0,1.05)$,  the system follows the dynamical transitions from traveling wave to 7-cluster chimera death state (7-CD) through complete synchronization and 7-cluster oscillation death states (7-OD).  On further increasing $\varepsilon_2$ beyond 1.05, the system attains 7-CD states through 7-OD states from traveling wave states.  When increasing the nonlocal coupling range to $r=0.4$, the system follows the same dynamical transition as observed in Fig. \ref{fig7}(a) except for the number of clusters which exist in the oscillation death and chimera death states.  Also, the synchronized region is enhanced for large values of nonlocal coupling range. One can note that for $r=0.1$, we can observe the 7- cluster oscillation death and chimera death states whereas in the case of $r=0.4$, one can note the 2-cluster oscillation death states and chimera death states.   More details on dynamical regimes are illustrated with space time plots and snapshots of the variables $x_j$ in Figs. \ref{fig7}(c)-(l).  Figures \ref{fig7}(c)-(d) depict the existence of traveling wave state for $(\varepsilon_2,\varepsilon_1)$ $\in$ $(1.36,0.25)$.  Then 7 coherent groups of upper and lower inhomogeneous steady states namely 7-cluster oscillation death state are illustrated with Figs. \ref{fig7}(e)-(f) for $(\varepsilon_2,\varepsilon_1)$ $\in$ $(1.36,1.00)$ with coupling range $r=0.1$.  In the strong coupling limit, the oscillators in the edges of the clusters randomly move between lower and upper inhomogeneous steady states and form 7-cluster chimera death state.  This is depicted in Figs. \ref{fig7}(g)-(h) for $(\varepsilon_2,\varepsilon_1)$ $\in$ $(1.36,2.50)$.  Interestingly, by increasing the coupling range, the number of clusters which exist in the oscillation death and chimera death states decreases.  That is for coupling range $r=0.4$, one can observe the 2 clusters of oscillation death state and 2 clusters of chimera death state which are illustrated in Figs. \ref{fig7}(i)-(j) for $(\varepsilon_2,\varepsilon_1)$ $\in$ $(1.7,1.1)$ and in Figs. \ref{fig7}(k)-(l) for $(\varepsilon_2,\varepsilon_1)$ $\in$ $(1.7,2.5)$, respectively.
\begin{figure*}[ht!]
\begin{center}
\resizebox{0.6\textwidth}{!}{\includegraphics{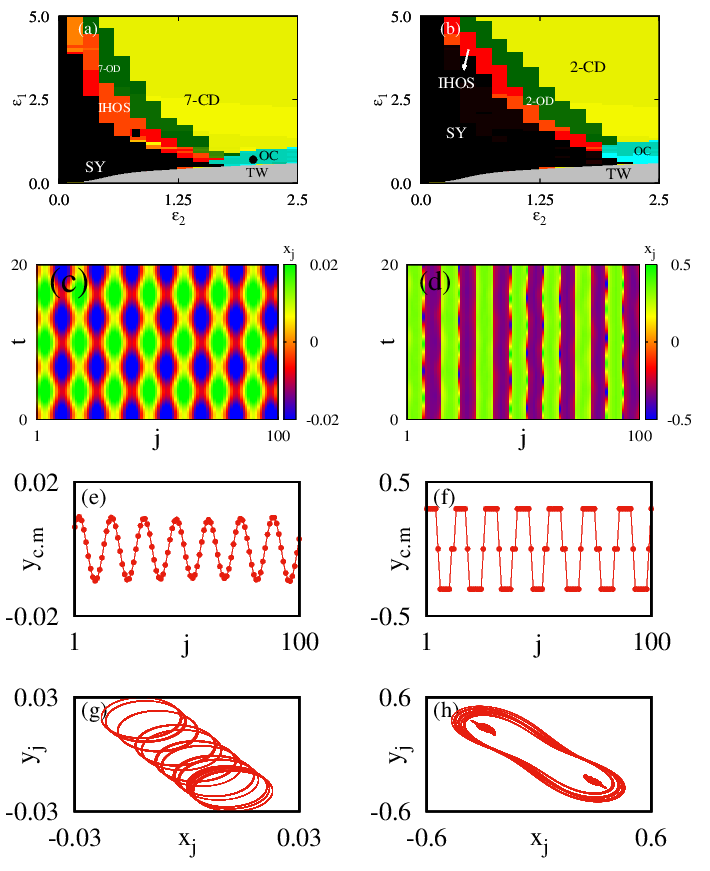}}\\
\end{center}
\caption{Two parameter phase diagram in the parametric space ($\varepsilon_2,\varepsilon_1$) in the absence of shear: (a) for nonlocal coupling range $r=0.1$ and (b) for nonlocal coupling range $r=0.4$.  The region SY is the synchronized region, $TW$ represents the traveling wave state, IHOS is the inhomogeneous oscillation state, OC represents the oscillatory cluster state, regions $7-OD$ and $7-CD$ represent 7-cluster oscillation death state and 7-cluster chimera death state, respectively and regions $2-OD$ and $2-CD$ represent 2-cluster oscillation death state and 2-cluster chimera death states, respectively.  Here $\bullet$ and $\blacksquare$ mark the parameter values corresponding to inhomogeneous oscillation state and oscillatory cluster state. Spatiotemporal plots depicting (c) inhomogeneous oscillation state for $(\varepsilon_2,\varepsilon_1)$ $\in$ $(0.70,1.4)$ and (d) cluster oscillatory state for $(\varepsilon_2,\varepsilon_1)$ $\in$ $(1.9,1.0)$ and their corresponding center of mass of each oscillator is shown in (e) and (f).  Phase portrait of the oscillators depicting (g) inhomogeneous oscillation state and (h) cluster oscillatory state. Other parameter values are $\omega=1$ and $c=3$.  Figures (c)-(h) depict the characteristic nature of the dynamical states with respect to nonlocal coupling range in the presence of shear.} 
\label{fig8}
\end{figure*} 
\par Next, we illustrate the impact of shear on dynamical transitions by plotting two parameter phase diagrams in the parametric space $(\varepsilon_2,\varepsilon_1)$ with fixed shear $c=3$ in Figs. \ref{fig8}(a) and \ref{fig8}(b) for $r=0.1$ and $r=0.4$, respectively.  By varying the value of $\varepsilon_1$ in the range of $\varepsilon_2$ $\in$ $(0.0,1.4)$, the system attains 7-CD from traveling wave state through synchronized state and inhomogeneous oscillatory state (IHOS) for $r=0.1$.  Here, the inhomogeneous oscillatory state represents the fact that the oscillators exhibit periodic oscillations with different centers of rotation as illustrated with the space-time plot in Fig. \ref{fig8}(c) for $(\varepsilon_2,\varepsilon_1)$ $\in$ $(0.70,1.4)$.  In order to illustrate the variation in the center of rotation, we have calculated the center of mass using the expression $y_{c.m}=\int_0^T\frac{y_j(t)dt}{T}$, $T=\frac{2\pi}{\omega}$ is the period of oscillation.  The center of mass corresponding to each oscillator is shown in Fig. \ref{fig8}(e).  Variation in the center of mass is also confirmed through the phase portrait of the oscillators in Fig. \ref{fig8}(g).  On further increasing the coupling interaction $\varepsilon_2>1.4$, the system attains the 7-CD state from traveling wave state through oscillatory cluster state.  In the oscillatory cluster state, one can observe homogeneous and inhomogeneous oscillatory clusters as depicted in Fig. \ref{fig8}(d) for $(\varepsilon_2,\varepsilon_1)$ $\in$ $(1.9,1.0)$.  Here the homogeneous oscillations (with large amplitude) are having the origin as a center of rotation while inhomogeneous oscillations (with small amplitude) are having nonzero value as a center of rotation which is illustrated with Fig. \ref{fig8}(f) and also verified through phase portrait of the oscillators given in Fig. \ref{fig8}(h).  From Fig. \ref{fig8}(b), one can observe the same dynamical transitions for $r=0.4$ as illustrated in Fig. \ref{fig8}(a) except for the number of clusters which exist in the oscillation death and chimera death states.  One can note that for $r=0.1$, we can observe the 7- cluster oscillation death and chimera death states whereas in the case of $r=0.4$, one can note the 2-cluster oscillation death states and chimera death states.  Thus we conclude that in the presence of shear under nonlocal coupling interaction, we can observe the symmetry breaking dynamical states namely the inhomogeneous oscillatory states and cluster oscillatory states.

\section{Conclusion}
\par In summary, we have investigated the interplay of shear with the asymmetric coupling involving attractive and repulsive interactions on dynamical states in globally coupled Stuart-Landau oscillators.  Symmetry breaking has been introduced in the form of asymmetric attractive and repulsive interactions.   In the absence of shear, the system follows a particular type of dynamical transition from desynchronized state to an oscillation death state through synchronized state.  By introducing a sufficient strength of shear with weak repulsive coupling interaction, the existence of solitary states, amplitude cluster states, synchronized states and oscillation death states has been noted along with appropriate transitions.  There occurs an increase of asymmetry in dynamical states by increasing the strength of repulsive coupling interaction which results in the emergence of imperfect solitary states and nontrivial oscillation death states.  Moreover, many dynamical transitions are observed with respect to the strength of the repulsive and attractive couplings. We have also verified that the analytical stability boundary for oscillation death state corroborates with the corresponding numerical boundary.  In addition, we have also investigated the impact of shear under nonlocal coupling interaction and found that the presence of shear leads to the existence of symmetry breaking dynamical states, namely inhomogeneous oscillation states and oscillatory cluster states.  Since asymmetry is more common in physical and biochemical systems, our study may provide a clue in controlling solitary states or chimera-like states under attractive and repulsive interactions.

 %%%%%%%%%%%%%%%%%%%%%%%%%%%%%%%%%%%%%%%%%%%%%%%%%%%%%%%%
\section*{Acknowledgments} 
The work of K. P has been supported by the UGC, Government of India through a Dr D S Kothari Post Doctoral Fellowship under Grant No. F.4-2/2006 (BSR) /PH/20-21/0197.  The work of V.K.C. forms part of the research projects sponsored by the DST-CRG Project under Grant No. CRG/2020/004353.  V.K.C. also wishes to thank DST, New Delhi for computational facilities under the DST-FIST Programme (Grant No. SR/FST/PS-1/2020/135) to the department of Physics. ML is supported by DST-SERB through a National Science Chair (Grant No. NSC/2020/000029).

\section*{Authors contribution statement}
All the authors contributed equally to the preparation of this manuscript.
\section*{Data availability statement}
The data that support the findings of this study are available within the article. 

\end{document}